# Ignition Delay Times of Kerosene (Jet-A)/Air Mixtures

V.P. Zhukov[*], V.A. Sechenov, A.Yu. Starikovskiy

Moscow Institute of Physics and Technology, Dolgoprudny, Russia

## *Abstract*

Ignition of Jet-A/air mixtures was studied behind reflected shock waves. Heating of shock tube at temperature of 150 $^0$C was used to prepare a homogeneous fuel mixture. Ignition delay times were measured from OH emission at 309 nm ($A^2\Sigma - X^2\Pi$) and from absorption of He-Ne laser radiation at 3.3922 μm. The conditions behind shock waves were calculated by one-dimensional shock wave theory from initial conditions $T_1$, $P_1$, mixture composition and incident shock wave velocity. The ignition delay times were obtained at two fixed pressures 10, 20 atm for lean, stoichiometric and rich mixtures (ϕ=0.5, 1, 2) at an overall temperature range of 1040-1380 K.

## *Introduction*

Ignition of large hydrocarbons determines the performance of internal combustion engines. Engines, which operate at high pressures, have higher efficiency and lower specific gravity. New up-to-date engines operate at higher pressures than the engines of older designs. Moreover, at high pressures the ignition occurs at lower temperature and so the $NO_x$ emission is smaller. Real fuels such as aviation kerosene and gasoline are a complicated mixture of hydrocarbons. The aviation kerosene contains very large hydrocarbons including, for example, hexadecane $C_{16}H_{34}$. Hexadecane as all large hydrocarbons has a very small pressure of saturated vapors at ambient temperature and boils at a very high temperature – 287 $^0$C. For large hydrocarbons the study of gas phase ignition requires heating of an experimental setup in order to prepare a homogeneous gas phase fuel mixture at desired pressure.

Ignition of large hydrocarbon has been studied extensively by various methods. Dagaut et al. [1] studied n-hexadecane combustion in jet-stirred reactor. The experiments were performed at 1 atm, over the temperature range 1000 to 1250 K and for equivalence ratios of 0.5, 1, and 1.5. Dagaut et al. measured a temperature dependence of mole fraction on outlet of reactor for molecular species (reactants, intermediates and final products). Lee et al. [2] measured minimum ignition energy by using laser spark ignition for hydrocarbons fuel in air at normal conditions. The minimum ignition energy of Jet-A fuel (commercial jet fuel) shows a nearly parabolic variation upon stoichiometric ratio of the mixture. The position of the minimum of the parabola depends upon temperature and pressure. Detonations of JP-10(cyclopentadiene)/air mixtures (pulse-detonation engine fuel) was studied by Austin and Shepherd [3].The experiments were performed in vapor-phase at 353 K over a range of equivalence ratios of 0.7 -1.4 and initial pressures of 20–130 kPa. For JP-10 mixtures cell widths of detonation wave were measured and founded to be comparable to those of propane and hexane mixtures. Ignition delay times and OH concentration time-histories for JP-10/$O_2$/Ar mixtures was measured behind reflected shock waves by Davidson et al. [4]. The experiments were performed over the temperature range of 1200-1700 K, pressure range of 1-9 atm, fuel concentrations of 0.2 and 0.4%, and stoichiometries of ϕ=0.5, 1.0 and 2.0. The fuel concentrations were measured in the shock tube using laser absorption at 3.39 μm. The ignition delay times were determinated using CH emission, and OH concentration histories were inferred from narrow-linewidth cw laser absorption measurements near 306 nm.

In works [5,6] ignition delay times for n-hexane/air and iso-octane/air mixtures were measured behind reflected shock waves. The experiments were performed at a pressure of

---

13-40 bar, a temperature of 700-1300 K, and equivalence ratio 0.5, 1, and 2. At temperatures below c.a. 1150 K, fuel-rich mixture ignites faster than fuel-lean mixtures. At 1150 K all mixtures ignite at approximately the same time and at temperatures above 1150 K fuel lean mixtures ignite faster than fuel rich mixtures.

In our pervious works we studied the ignition of lean $\phi=0.5$ hydrocarbon/air mixtures [7,8]. The ignition delay times were measured behind reflected shock waves for lean methane/air, propane/air and n-hexane/air mixtures at wide pressure range of 2-500 atm and wide overall temperature range of 800-1700 K. The experimental data demonstrates a good agreement with experimental results of other papers [9,10]. The ignition delay times calculated for methane by kinetic model [11] and for n-hexane by model [12] coincide with our experimental data in the range of uncertainties. Developed and approbated methods are used in the current work for measurements and interpretations of experimental data on ignition of Jet-A/air mixtures.

## *Experimental Setup*

All experimental work of the current study was performed at the Physics of Nonequilibrium Systems Laboratory at Moscow Institute of Physics and Technology. All measurements presented were obtained in a helium-driven preheated high-pressure shock tube.

A shock tube has some advantages for investigation of ignition processes in a wide range of parameters: 1) the influence of high pressures and temperatures on tube walls and windows lasts for a short time interval; 2) required thermodynamic parameters of test gas behind the incident and reflected shock waves can be obtained by a variation of initial conditions: pressure, temperature, composition of mixture, and driver gas pressure; 3) the gas behind the reflected shock wave is stagnate and uniform in space.

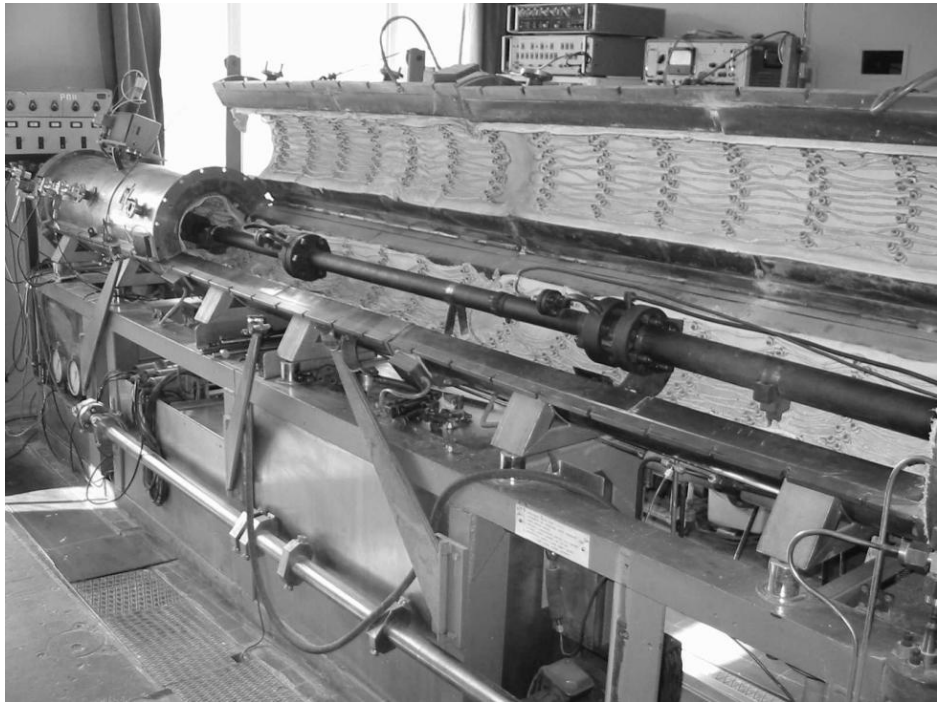

**Figure 1. Shock tube placed into heater.**

## Shock Tube

A preheated shock tube with 45 mm inner diameter is made of stainless steel with a 0.7 m driver section and at 3.2 m test section. The test and the driver sections are separated with a two-diaphragm auxiliary chamber, which is meant for a diaphragm rupture at a fixed driver gas pressure. The shock tube is placed inside a heater, which allows the heating of the tube up to temperature of 900 K. Eight thermocouples are placed along the tube at different points for a temperature registration. The heater is divided into three independent sections. The overall view of the shock tube and the heater is shown in figure 1. Fixed temperature can be maintained along the tube with the accuracy of 5 K with the aid of a special automatic system.

A mixing tank is made of stainless steel and placed inside the heater. The volume of the mixing tank is smaller by factor of 3 than the volume the shock tube. A tilting of the tank about horizontal axis can be varied within the range of ±1.5 degrees. To stir the mixture a brass ball of 43 mm in diameter is placed inside the mixing tank and can be moved in the tank. If it is needed the pneumatic pump can raise the pressure in the driver section up to 1000 atm. A vacuum sealing of the shock tube is made of annealed copper and tube windows are made of sapphire. These allow carrying out study of non-volatile liquids behind the incident and reflected shock waves at high pressures. A scheme of the experimental setup is shown on Figure 2.

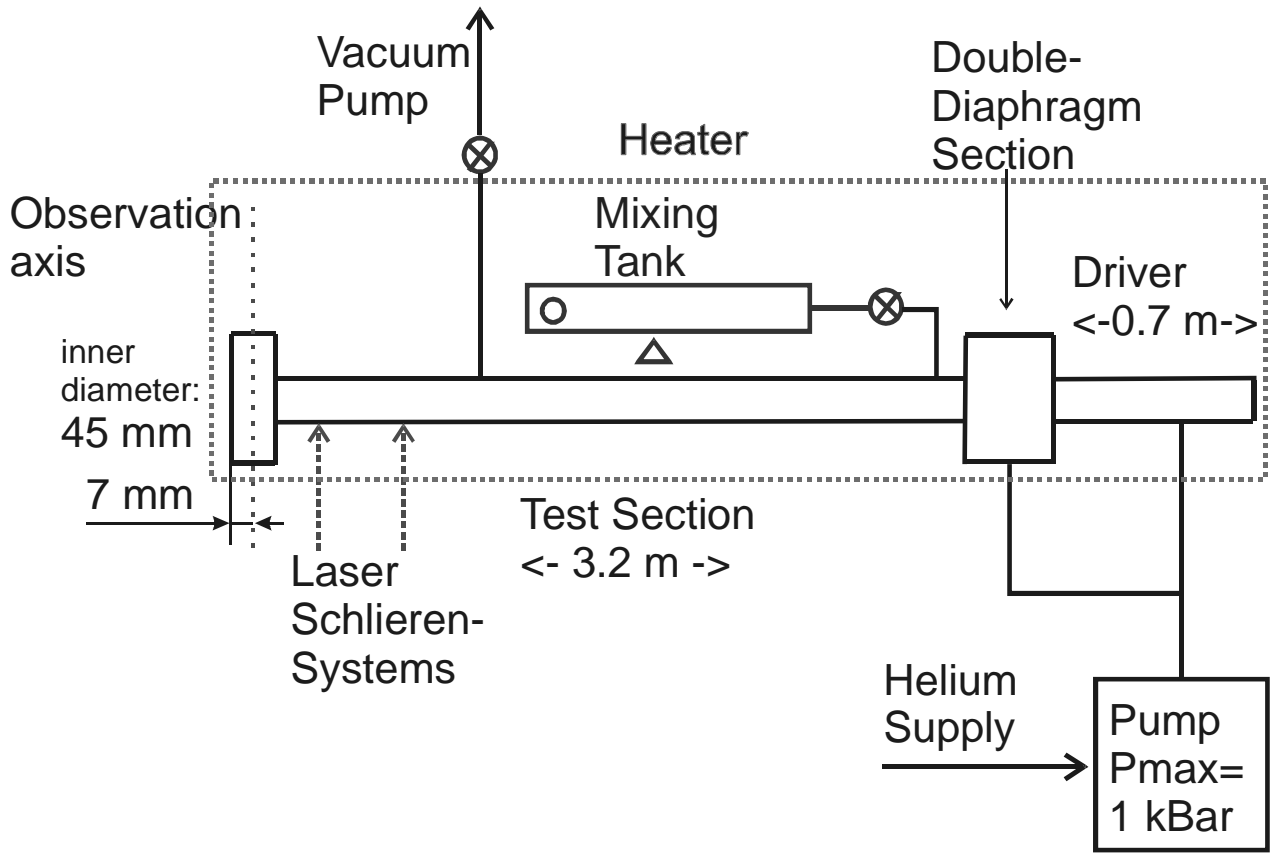

**Figure 2. Scheme of experimental setup.**

## Diagnostic system

The ignition was observed through the sidewall at distance of 7 mm away from the end wall. An emission UV and absorption IR spectroscopy is used to measure ignition delay time.

The UV diagnostic system consists of a grating monochromator and a photo-electronic multiplier. This technique allows us to measure the time profile of electron-excited OH emission ($A^2\Sigma - X^2\Pi$) at wavelength of 309 nm.

In addition to this method, an absorption diagnostic system is used. It consists of an IR He-Ne laser ($\lambda=3.3922$, which corresponds to the absorption in the asymmetric $\nu_3$ mode of $CH_3$ group) and a PbSe – photo-resistor. The absorption profile changes at the time moment of incident and reflected shock waves passing through the observation axis. After the ignition, a considerable decrease in the absorption of the laser radiation was observed simultaneously with the emission peak at 309 nm.

## *Experimental Measurements*

Before each experiment the shock tube and the mixing tank was cleaned to remove kerosene and soot.

The mixing tank was placed inside the heater and connected with test section of the shock tube. The test section and manifold were vacuumed to $3\text{-}4\cdot10^{-2}$ torr. Then the heat was turned on. The tube was heated to an operating temperature of 150 $^0$C uniformly with the aid of special automatic system. After the heating a forced stirring of fuel mixture was carried out.

After evacuating and heating the vacuum manifold was blocked and a mixer valve was opened. Measurements of initial pressure $P_1$ of the test gas were fulfilled using a buffer gas (air). The buffer gas filled the manifolds outside the heater at a pressure that was close to a pressure of the test gas.

## **Mixture preparation**

A fuel mixture was consisted of aviation fuel Jet-A and dried atmospheric air. For the mixture preparation the procedure was as follows:
1. the mixing tank was evacuated up to a residual gas pressure of $3\text{-}4\cdot10^{-2}$ torr;
2. the mixing tank was filled with dried air up to atmospheric pressure;
3. a fixed amount of Jet-A was poured into the mixing tank;
4. the mixing tank was filled with dried air up to a required pressure.

A quantity of the fuel charge was measured by volume from 0.06 ml to 0.6 ml. The fuel volume was measured with accuracy better than 0.02 ml. To determine a mixture composition we recalculated a charge volume to a charge mass. The measured density of used Jet-A sample was $(0.80\pm0.03)$ g/ml at 26 $^0$C. This value lies within the range of specification ASTM D 1655 (0.775-0.840 g/ml at 15 $^0$C).

To obtain the homogenous fuel mixture we forced stirred the test mixture. The stirring was run at temperature of 150 $^0$C directly before the filling of the test section. We performed separate experiments with various amounts of the stirring. The experiments shown that the ignition delay times depend on the stirring at amounts of the stirring of 8-15 and doesn't depend on the stirring at amounts of the stirring of 20-30. We performed 40-50 amounts of stirring at the measurements of the ignition delay times for Jet-A/air mixtures.

The fuel Jet-A is a mixture of a large number of hydrocarbons (see Figure 3). It is difficult to determine the composition of fuel Jet-A. The chromatogram of our Jet-A sample coincides with the typical chromatogram provided in [**Ошибка! Источник ссылки не найден.**] (see Figure 3 and Figure 4). We used data on Jet-A composition from the work [**Ошибка! Источник ссылки не найден.**]. To calculate the fuel mixture composition and stoichiometry we supposed that an approximate formula for Jet-A was $C_{11}H_{21}$.

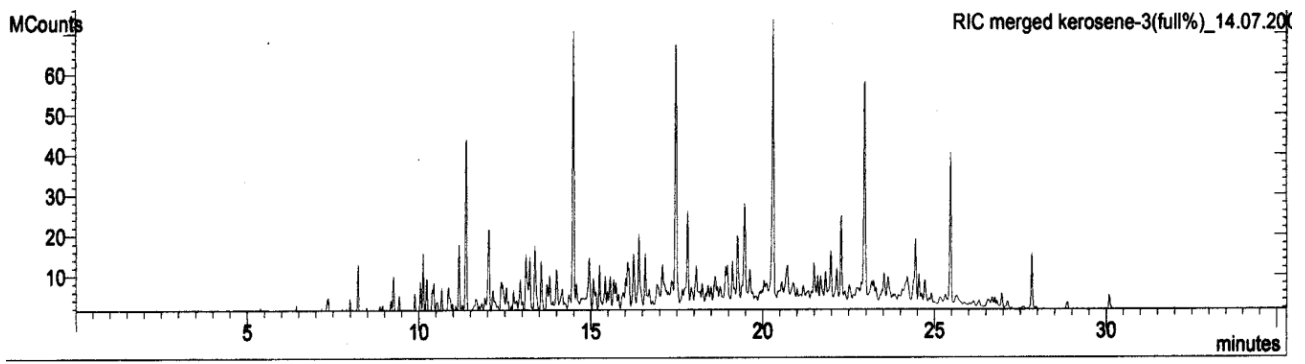

**Figure 3. Chromatogram of investigated Jet-A sample.**

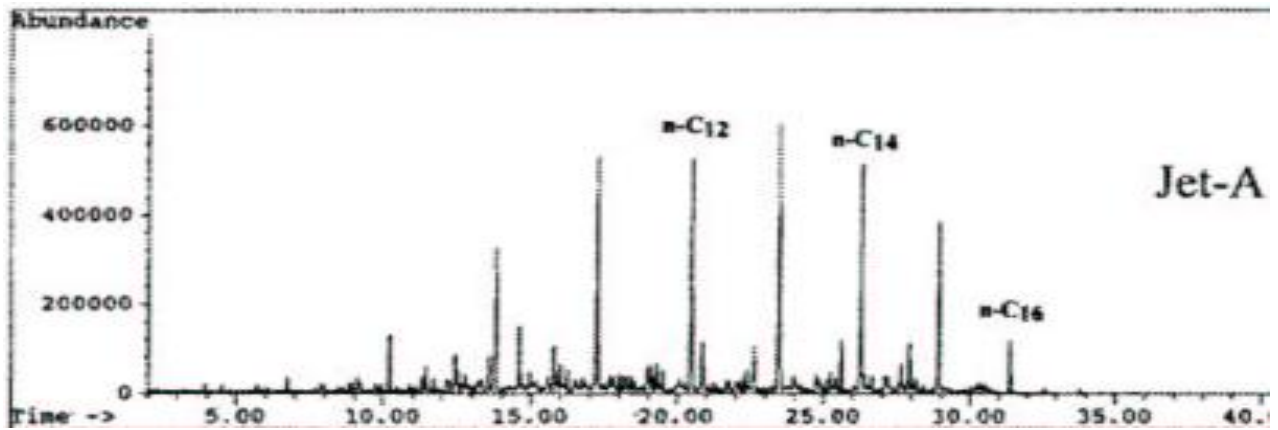

**Figure 4. Typical chromatogram of fuel Jet-A from the work** [Ошибка! Источник ссылки не найден.]**.**

## Ignition Delay Time Definition

The ignition delay times were determined behind reflected shock waves. The papameters of fuel mixture behind the shock waves were calculated by one-dimensional shock wave theory assuming full relaxation and "frozen" chemistry and using the measured incident shock wave velocity extrapolated to the end wall of the shock tube. The measured ignition times are greater than 12 μs, that is long enough to reach the full relaxation in fuel/air mixture, in which kerosene fraction is greater than 0.6%. The rate of chemical reactions depends on temperature exponentially therefore the kinetics behind incident shock waves is slower than that behind reflected shock waves. In our experiments the ignition delay times were much more than 1 ms at the hottest conditions behind incident shock waves. This time is much more than time between incident and reflected shock waves ($\Delta\tau\sim30$ μs). Moreover, the assumption about "frozen" chemistry was confirmed by experimental observations. We didn't observe the emission at 309 nm behind the incident shock wave. The temperature of the end wall was taken as the initial temperature of test gas. The source of thermodynamic properties was [16] for Jet-A and [17] for air components. In the calculations the air was approximated by the following composition: $N_2:O_2:Ar=78.12:20.95:0.93$.

The ignition delay time was defined as the time difference between the reflected shock wave appearance in the observation section and the emission peak at $\lambda=309$ nm at the same section (see Figure 5). On observation axis the ignition promoted by the earlier ignition, which occurs near the end wall. The ignition delay time measured through the sidewall shorter than the one measured through the end wall. The ignition delay times of fuel/air mixtures were measured at the end plate and apart from the end plate simultaneously in [13]. The conditions of this experiment (shock tube parameters, pressure, temperature, and heat release behind the shock waves) are very similar to the conditions of current study. The difference between the endwall and the sidewall measurement was ~20 μs for a distance of 20 mm between the observation points. We increased all the measured ignition delay times by 7 μs.

Behind the reflected shock wave the temperature increases because of the boundary layer effect [15]. The rate of temperature increase depends on the initial gas pressure and specific heat ratio. For low pressure limit of this work ($P_1$~0.2 atm) the rate may be estimated as $dT/dt < 0.05$ K/μs. At ignition delay times of <400 μs the temperature variation due to non-ideal effects is smaller than the measurement uncertainties in the temperature.

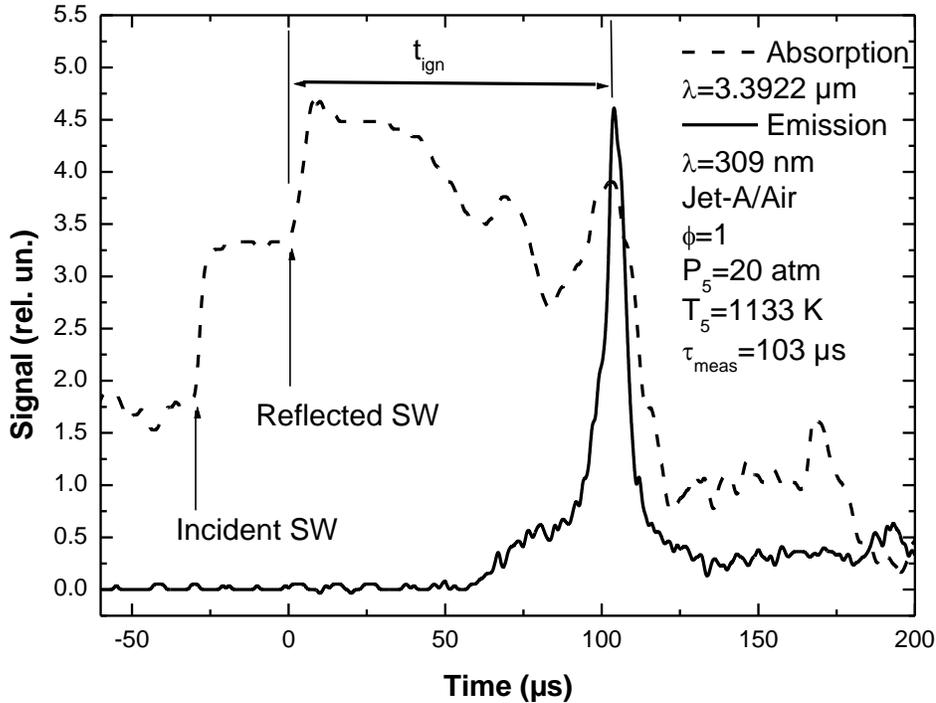

**Figure 5. Definition of ignition delay time.**

## Uncertainties in Measurements

The uncertainties in ignition delay times can be subdivided into two general classes – parametric uncertainties and measurement uncertainties. The parametric uncertainties are defined as those due to uncertainties in the relevant test parameters (i.e., pressure, temperature, mixture composition), while measurement uncertainties are associated with the inaccuracies inherent in measuring and quantifying the ignition time.

The measurement uncertainty consists of three components:
1. the definition of emission peak,
2. an uncertainty in the definition of reflected shock wave passing time,
3. an uncertainty associated with the correction of the ignition delay times measured through the sidewall.

The measurement uncertainties amounted to 3-7 μs in our experiment.

The ignition time of hydrocarbons depends on the temperature strongly. The uncertainties in the temperature behind the reflected shock wave leads to uncertainties in the ignition delay time.

The uncertainties in the temperature behind the reflected shock wave originates from 1% inaccuracy of measurements of shock wave velocity and 1% inaccuracy of measurements of test gas temperature $T_1$. The total uncertainties in the temperature behind the reflected shock wave amount to 1.5%. Thus these uncertainties correspond to 8 μs error for 10-100 μs the ignition delay time, 25 μs error for 100-200 μs the ignition time and 60 μs for the 200-400 μs.

## *Discussion*

The ignition delay times of Jet-A follows the Arrhenius law at temperatures of 1000-1400 K. The obtained experimental data can be approximated by a correlation expression (see Figure 6). This approximation allows us to compare the obtained results with results of other works. The activation energy equals to 29.2±1.3 kcal/mol for Jet-A/air mixtures. This value is slightly less than that for saturated hydrocarbons ~47 kcal/mol [18]. The difference can be explained by high contents of non-paraffin species (~35%).

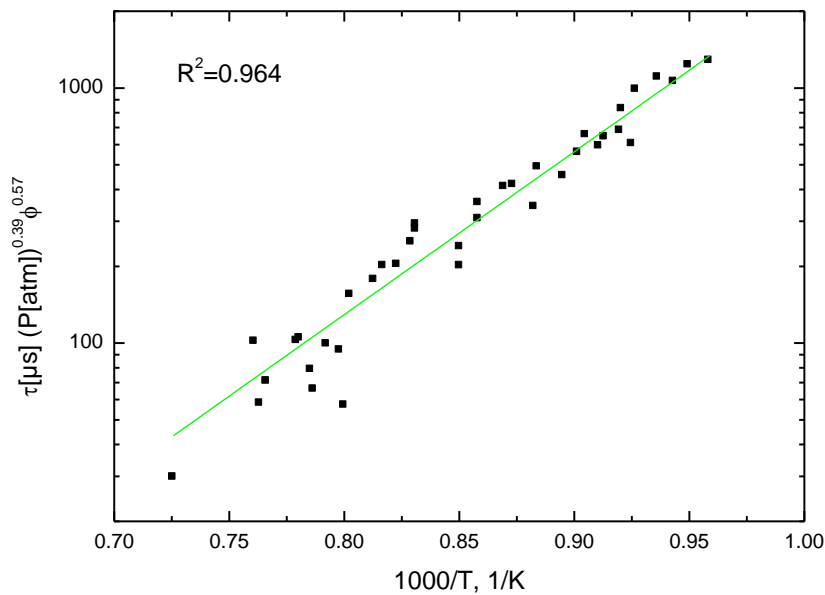

**Figure 6. Correlation expression for ignition delay times of Jet-A/air mixtures.**

The ignition delay times of Jet-A/air mixtures decrease with the pressure raise as typical for hydrocarbons and with increasing of equivalence ratio. The same dependence on the equivalence ratio is observed for n-heptane and iso-octane [5,6]. The ignition delay times of low hydrocarbons increase with the stoichiometry. For low hydrocarbons the fuel inhibits the ignition. This is due to the nature of the chain-branching process. For small hydrocarbons, which ignite at a relatively high temperature, it occurs because of the reaction of hydrogen atom with molecular oxygen produces a hydroxyl radical and an oxygen atom. For large hydrocarbons at temperature about 1000 K the chain-branching generates by hydrocarbon hydroperoxide species, whose production is directly proportional to the fuel concentration.

## *Conclusions*

The ignition delay times were measured at the two fixed pressures 10, 20 atm for the lean, stoichiometric and rich mixtures (φ=0.5, 1, 2) at the overall temperature range of 1040-1380 K.

It was shown that the ignition delay times monotone decrease with the pressure and temperature raise. The dependence of ignition delay times on equivalence ratio allows to suppose that chain-branching mechanism in kerosene-air mixtures is the same as in large alkane-air mixtures – with hydropexide species production.

The experimental data presented on tables and Arrhenius plots in the appendix-A. The obtained experimental data can applied directly in commercial applications and can used for a verification of kinetic mechanisms.

## *Appendix A – Ignition Delay Time Data*

**Table 1 Jet-A/Air, P=20 atm, ϕ=1**

| Temperature [K] | Pressure [atm] | Jet-A [% vol.] | Air [% vol.] | ϕ | Ignition Delay Time [µs] |
|---|---|---|---|---|---|
| 1176 | 20.4 | 1.27 | 98.73 | 1.00 | 76 |
| 1133 | 20.0 | 1.26 | 98.74 | 0.99 | 110 |
| 1060 | 20.7 | 1.25 | 98.75 | 0.98 | 336 |
| 1117 | 20.9 | 1.26 | 98.74 | 0.99 | 143 |
| 1273 | 20.7 | 1.26 | 98.74 | 0.99 | 25 |
| 1053 | 18.7 | 1.26 | 98.74 | 0.99 | 406 |
| 1105 | 22.3 | 1.26 | 98.74 | 0.99 | 202 |

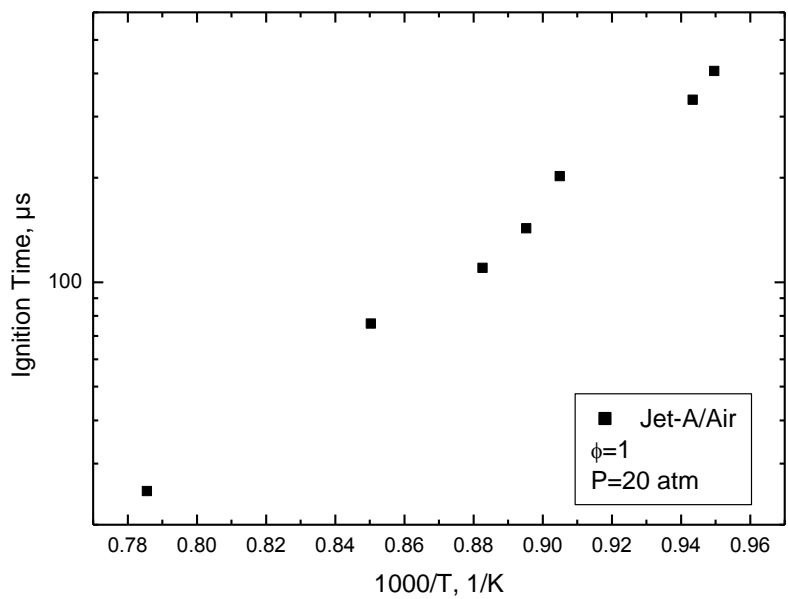

**Table 2 Jet-A/Air, P=20 atm, $\phi$=0.5**

| Temperature [K] | Pressure [atm] | Jet-A [% vol.] | Air [% vol.] | $\phi$ | Ignition Delay Time [μs] |
|---|---|---|---|---|---|
| 1310 | 21.7 | 0.63 | 99.37 | 0.49 | 27 |
| 1253 | 21.3 | 0.63 | 99.37 | 0.49 | 44 |
| 1098 | 17.1 | 0.63 | 99.37 | 0.49 | 302 |
| 1165 | 18.0 | 0.63 | 99.37 | 0.49 | 170 |
| 1145 | 20.2 | 0.63 | 99.37 | 0.49 | 209 |
| 1215 | 20.8 | 0.63 | 99.37 | 0.49 | 96 |
| 1095 | 19.1 | 0.63 | 99.37 | 0.49 | 314 |
| 1165 | 16.1 | 0.63 | 99.37 | 0.49 | 160 |
| 1230 | 18.3 | 0.63 | 99.37 | 0.49 | 88 |
| 1262 | 20.6 | 0.63 | 99.37 | 0.49 | 47 |

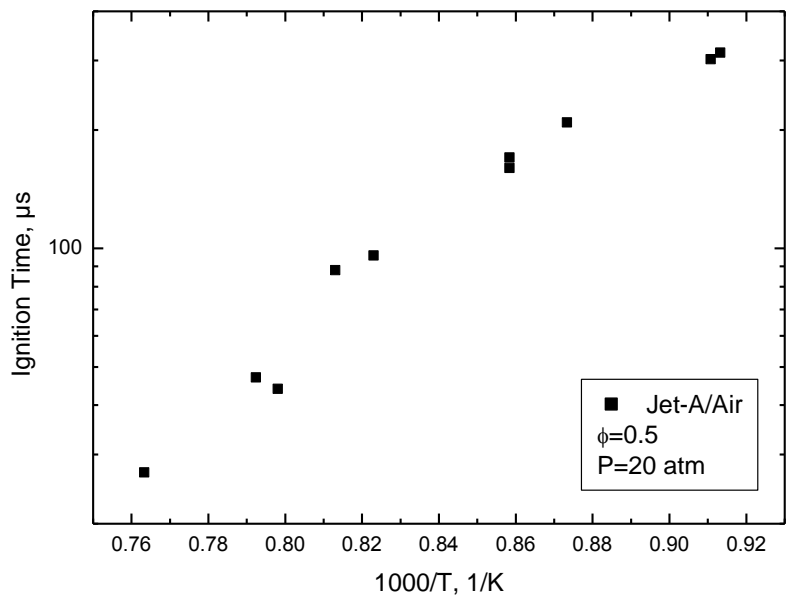

**Table 3 Jet-A/Air, P=20 atm, $\phi$=2**

| Temperature [K] | Pressure [atm] | Jet-A [% vol.] | Air [% vol.] | $\phi$ | Ignition Delay Time [μs] |
|---|---|---|---|---|---|
| 1043 | 22.4 | 2.54 | 97.46 | 2.02 | 263 |
| 1081 | 21.9 | 2.54 | 97.46 | 2.02 | 125 |
| 1068 | 18.5 | 2.54 | 97.46 | 2.02 | 243 |
| 1176 | 20.0 | 2.54 | 97.46 | 2.02 | 43 |
| 1271 | 17.0 | 2.54 | 97.46 | 2.02 | 15 |
| 1109 | 27.1 | 2.54 | 97.46 | 2.02 | 78[*] |
| 1145 | 29.4 | 2.54 | 97.46 | 2.02 | 66[*] |

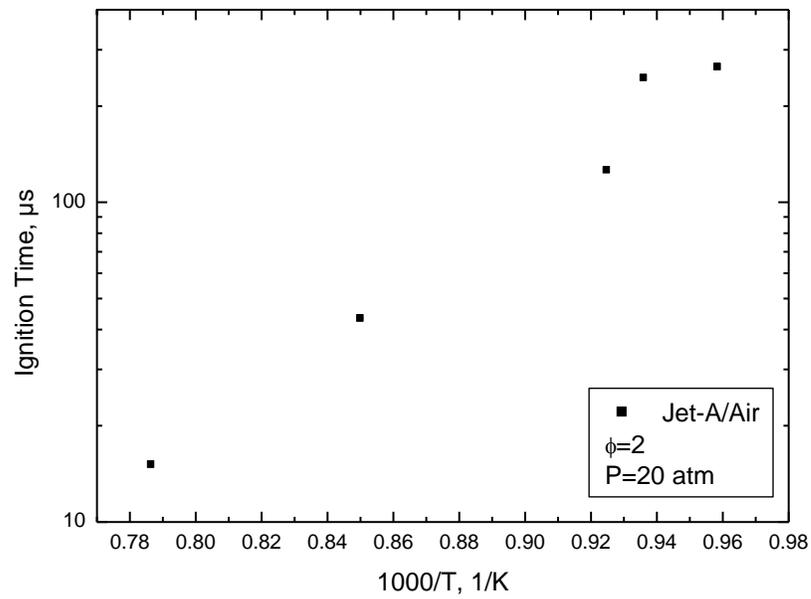

[*] Not included into the plot.

**Table 4 Jet-A/Air, P=10 atm, φ=1**

| Temperature [K] | Pressure [atm] | Jet-A [% vol.] | Air [% vol.] | φ | Ignition Delay Time [μs] |
|---|---|---|---|---|---|
| 1246 | 10.6 | 1.27 | 98.73 | 1.0 | 63 |
| 1283 | 11.0 | 1.27 | 98.73 | 1.0 | 41 |
| 1281 | 9.72 | 1.27 | 98.73 | 1.0 | 44 |
| 1206 | 10.7 | 1.27 | 98.73 | 1.0 | 101 |
| 1109 | 10.3 | 1.27 | 98.73 | 1.0 | 230 |
| 1079 | 11.0 | 1.27 | 98.73 | 1.0 | 396 |

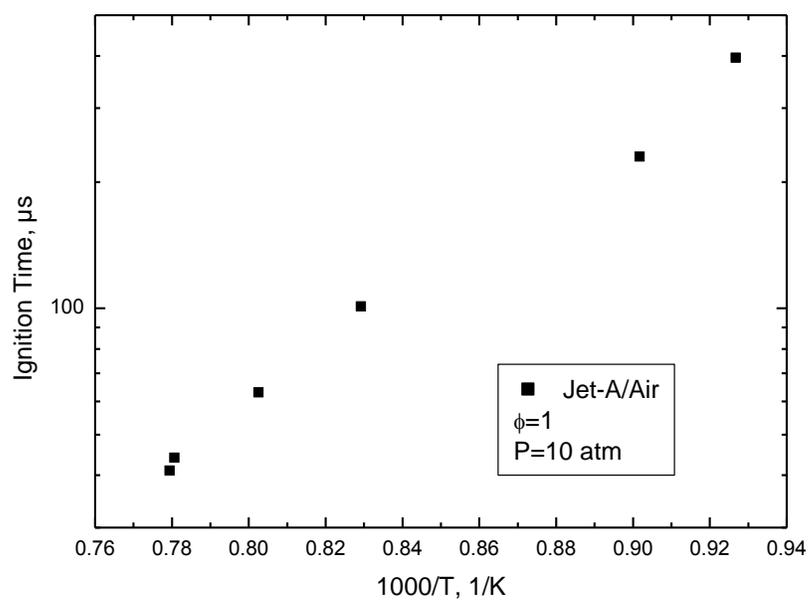

**Table 5 Jet-A/Air, P=10 atm, $\phi$=0.5**

| Temperature [K] | Pressure [atm] | Jet-A [% vol.] | Air [% vol.] | $\phi$ | Ignition Delay Time [µs] |
|---|---|---|---|---|---|
| 1203 | 10.0 | 0.64 | 99.36 | 0.5 | 173 |
| 1224 | 11.2 | 0.64 | 99.36 | 0.5 | 119 |
| 1150 | 11.9 | 0.64 | 99.36 | 0.5 | 237 |
| 1087 | 10.3 | 0.64 | 99.36 | 0.5 | 416 |
| 1378 | 10.6 | 0.64 | 99.36 | 0.5 | 18 |
| 1305 | 9.3 | 0.64 | 99.36 | 0.5 | 45 |

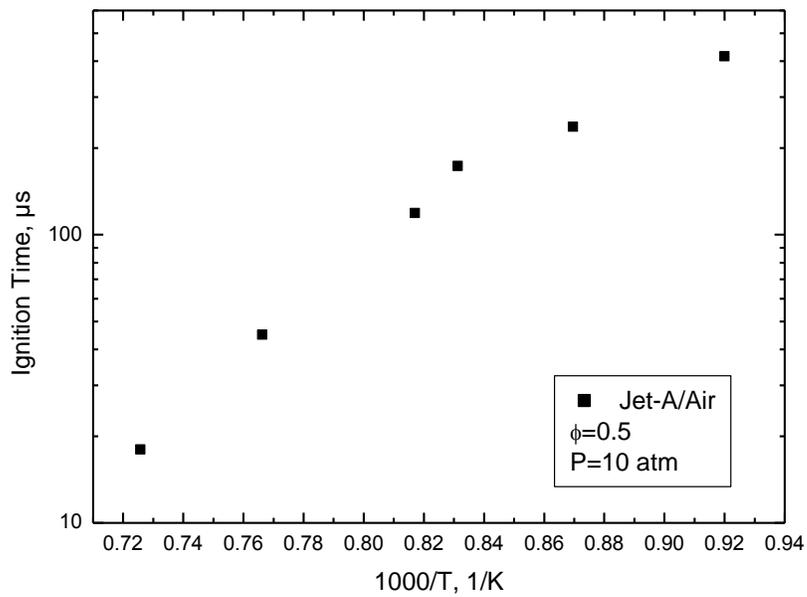

**Table 6. Jet-A/Air, P=10 atm, φ=2**

| Temperature [K] | Pressure [atm] | Jet-A [% vol.] | Air [% vol.] | φ | Ignition Delay Time [μs] |
|---|---|---|---|---|---|
| 1131 | 11.3 | 2.51 | 97.49 | 2.0 | 131 |
| 1314 | 14.0 | 2.51 | 97.49 | 2.0 | 20 |
| 1086 | 12.4 | 2.51 | 97.49 | 2.0 | 214 |
| 1203 | 11.4 | 2.51 | 97.49 | 2.0 | 78 |

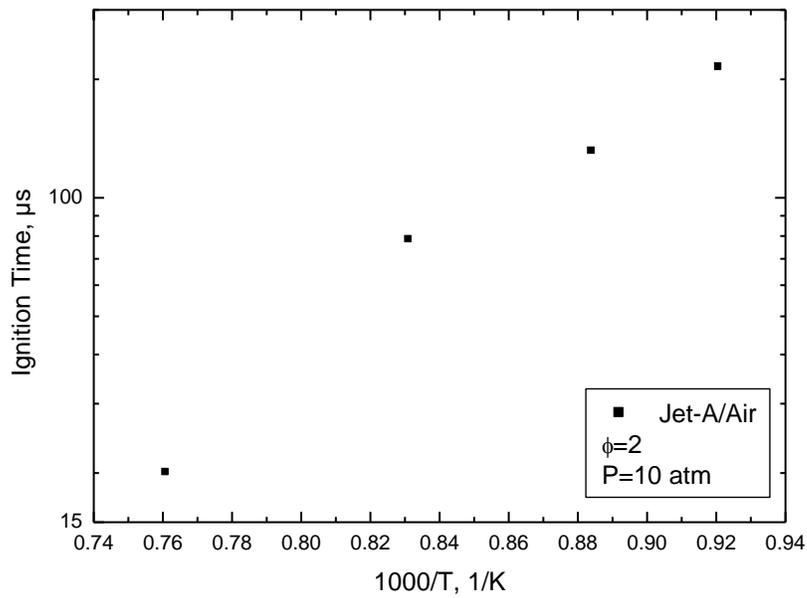